\documentclass[sigconf,nonacm]{acmart}
\AtBeginDocument{%
  }

\usepackage{enumitem}
\usepackage{multirow}
\usepackage{listings}
\usepackage{makecell}
\usepackage{pifont}
\newcommand{\cmark}{\textcolor{green}{\ding{51}}}
\newcommand{\xmark}{\textcolor{red}{\ding{55}}}
\usepackage{booktabs}
\usepackage{subcaption}
\usepackage[listings]{tcolorbox}
\usepackage{fancyvrb}
\usepackage{xcolor}
\tcbuselibrary{skins,breakable}
\definecolor{delim}{RGB}{20,105,176}
\definecolor{numb}{RGB}{106, 109, 32}
\definecolor{string}{rgb}{0.64,0.08,0.08}
\usepackage{xcolor}
\usepackage[dvipsnames]{xcolor}
\usepackage{listings}
\definecolor{jsonKey}{RGB}{0, 92, 197}       
\definecolor{jsonString}{RGB}{163, 21, 21}   
\definecolor{jsonNumber}{RGB}{9, 134, 88}    
\definecolor{jsonPunct}{RGB}{60, 60, 60}     
\definecolor{jsonBg}{RGB}{248, 248, 248}     
\definecolor{aesBlue}{RGB}{64,120,192}    
\definecolor{aesTeal}{RGB}{22,147,165}    
\definecolor{aesCoral}{RGB}{200,72,72}    
\definecolor{aesPurple}{RGB}{138,99,181}  
\definecolor{aesAmber}{RGB}{200,130,30}   
\definecolor{aesBg}{RGB}{250,250,248}     
\lstdefinelanguage{json}{
    basicstyle=\ttfamily\small,
    numbers=none,
    frame=single,
    rulecolor=\color{gray!40},
    backgroundcolor=\color{jsonBg},
    showstringspaces=false,
    breaklines=true,
    breakatwhitespace=true,
    sensitive=true,
    morestring=[b]",
    literate=
     *{0}{{{\color{jsonNumber}0}}}{1}
      {1}{{{\color{jsonNumber}1}}}{1}
      {2}{{{\color{jsonNumber}2}}}{1}
      {3}{{{\color{jsonNumber}3}}}{1}
      {4}{{{\color{jsonNumber}4}}}{1}
      {5}{{{\color{jsonNumber}5}}}{1}
      {6}{{{\color{jsonNumber}6}}}{1}
      {7}{{{\color{jsonNumber}7}}}{1}
      {8}{{{\color{jsonNumber}8}}}{1}
      {9}{{{\color{jsonNumber}9}}}{1}
      {:}{{{\color{jsonKey}\bfseries:}}}{1}
      {,}{{{\color{jsonPunct}\bfseries,}}}{1}
      {[}{{{\color{violet}\bfseries[}}}{1}
      {]}{{{\color{violet}\bfseries]}}}{1}, 
    string=[s]{"}{"},
    stringstyle=\color{jsonString},
    comment=[l]{//},
    commentstyle=\color{gray}\itshape,
}
\begin{document}

\title{LLM-based Hardware Development with Hierarchical IRs and End-to-End Multi-Agent Workflow}


\author{Chenyang Yin}
\email{ycy@u.nus.edu}
\affiliation{
  \institution{National University of Singapore}
  \country{Singapore}
}

\author{Agasthi Haputhanthri}
\email{agasthi31@nus.edu.sg}
\affiliation{
  \institution{National University of Singapore}
  \country{Singapore}
}

\author{Aditya Anirudh Jonnalagadda}
\email{aditya.anirudh@u.nus.edu}
\affiliation{
  \institution{National University of Singapore}
  \country{Singapore}
}

\author{Zhenyu Bai}
\email{zhenyu.bai@nus.edu.sg}
\affiliation{
  \institution{National University of Singapore}
  \country{Singapore}
}

\author{Yuanming Song}
\email{yuanming@mail.sdu.edu.cn}
\affiliation{
  \institution{Shandong University}
  \country{China}
}

\author{Saranyu Chattopadhyay}
\email{saranyuc@stanford.edu}
\affiliation{
  \institution{Stanford University}
  \country{USA}
}

\author{Mohammad Fadiheh}
\email{fadiheh@stanford.edu}
\affiliation{
  \institution{LUBIS EDA}
}

\author{Tom Zelazny}
\email{tzelazny@stanford.edu}
\affiliation{
  \institution{Stanford University}
  \country{USA}
}

\author{Subhasish Mitra}
\email{subh@stanford.edu}
\affiliation{
  \institution{Stanford University}
  \country{USA}
}

\author{Tulika Mitra}
\email{tulika@comp.nus.edu.sg}
\affiliation{
  \institution{National University of Singapore}
  \country{Singapore}
}


\newcommand{\customauthorblock}{%
  \centering

  {\large
  Chenyang Yin\textsuperscript{1}
  \qquad
  Agasthi Haputhanthri\textsuperscript{1}
  \qquad
  Aditya Anirudh Jonnalagadda\textsuperscript{1}
  \\[0.35em]
  Zhenyu Bai\textsuperscript{1}
  \qquad
  Yuanming Song\textsuperscript{2}
  \qquad
  Saranyu Chattopadhyay\textsuperscript{3}
  \\[0.35em]
  Mohammad Fadiheh\textsuperscript{4}
  \qquad
  Tom Zelazny\textsuperscript{3}
  \qquad
  Subhasish Mitra\textsuperscript{3}
  \qquad
  Tulika Mitra\textsuperscript{1}
  \par}

  \vspace{0.8em}

  {\normalsize
  \textsuperscript{1}National University of Singapore
  \qquad
  \textsuperscript{2}Shandong University
  \qquad
  \textsuperscript{3}Stanford University
  \qquad
  \textsuperscript{4}LUBIS EDA
  \par}
}

\makeatletter
\renewcommand{\@mkauthors@iii}{%
  \global\setbox\mktitle@bx=\vbox{%
    \noindent
    \unvbox\mktitle@bx
    \par
    \medskip
    \customauthorblock
    \par
    \bigskip
  }%
}
\makeatother

\renewcommand{\shortauthors}{Yin et al.}

\begin{abstract}
Large language models (LLMs) are increasingly used in software development, but their use in complex hardware design remains limited. This gap stems from both the scarcity of public hardware training data and the fundamentally different methodologies used in hardware design. In particular, applying LLMs to hardware requires more than direct RTL generation: the model must understand module boundaries, inter-module connections, and verification requirements. For nontrivial designs, directly generating complete RTL using LLM is rarely effective. 

In this paper, we present an LLM-based hardware development framework with hierarchical intermediate representations (IRs) and an end-to-end multi-agent workflow. The core idea is to provide an abstraction of hardware design to LLMs through two structured IRs: Architectural Sketch, which captures module topology and interconnection, and Operational Specification, which defines per-module functionality and interfaces. Our framework uses these IRs to decompose a complex design into sub-modules, specify the per-block functionality, and derive how each module should be tested and verified. In order to fix the intractable errors in complex designs, we incorporate a multi-agent debug loop in the framework, allowing agents to get the error feedback and control the debug details such as the signals to be probed for simulation. 

We evaluate our framework on Verilog-Eval benchmark, achieving a pass@5 rate of 95.5\%, which surpasses current state-of-the-art LLM generation frameworks. To better assess performance on complex, realistic designs, we introduce a new case study spanning applications from general-purpose processors to digital signal processing systems. Experimental results indicate that such complex designs exceed the capabilities of existing approaches, whereas our framework is the only one capable of producing functional end-to-end design. Our generated RTL follows all industry-standard design rules, is lint-clean, functionally correct and fully synthesizable. Within the case study, each case averages 780 lines of RTL and synthesizes to a netlist of roughly 1.4 million cells, with RTL generation taking around 800 seconds and consuming approximately 82,000 tokens per design.

\end{abstract}

\maketitle

\section{Introduction}
As large language models (LLMs) continue advancing, cutting-edge models such as GPT-5.4\cite{openai2025gpt5}, Claude 4.6\cite{anthropic2025claude}, and Gemini 3.1\cite{google2025gemini} have demonstrated remarkable capabilities in assisting software engineering tasks \cite{LLMsoft2}\cite{LLMsoft1}. Leveraging LLM-powered programming tools like GitHub Copilot\cite{github2023copilot} and Cursor\cite{cursor2024}, software engineers have significantly reduced manual effort and enhanced productivity. More recently, emerging agentic tools, like Claude Code\cite{claudecode} and Openclaw\cite{openclaw}, have substantially lowered barriers to software development and improved developer productivity.

This on-going LLM trend presents a precious opportunity for the hardware industry. With a high barrier to entry, the hardware design process remains highly challenging and time-consuming even for advanced engineers. This situation highlights the significant potential of leveraging LLMs to assist hardware development. 
Existing studies such as \cite{chipnemo,verigen,liu2024rtlcoder,cui2024origen,pei2024betterv,wei2025vericoder,zhao2025mage,rtl++,sami2024eda,reasoningv} have demonstrated the effectiveness of LLM-aided hardware design by achieving strong results on relevant benchmarks such as Verilog-Eval\cite{liu2023verilogeval} and RTLLM\cite{lu2024rtllm}. However, these works are limited by their main focus on relatively small, single-module hardware designs.  
In contrast, real-world hardware development tasks—where LLM assistance is most needed—are typically more complex. They involve multiple interacting modules, sophisticated algorithms, and require users to develop their own testbenches rather than those provided in the benchmark. 
This disconnect highlights a clear gap between current research and the demands of real-world hardware design.


Real-world hardware development typically begins with defining the top-level architecture, which establishes the system organization. Individual functional blocks are then designed incrementally, with each block verified using its own testbench. Once all sub-modules are validated, a top-module is constructed to integrate them. The system then undergoes system-level verification using a top-level testbench, ultimately yielding an implementation-ready hardware design. 
While existing research places much attention on assisting the RTL design stage with LLMs, the verification stage—providing critical feedback needed for ensuring functional correctness—is in fact highly time-consuming, with a median share of 50–60\% of the overall ASIC design time~\cite{foster2022wilson}.
The end-to-end workflow requires substantial time and human effort, and using LLMs only for RTL generation fails to fully exploit their potential.

Motivated by this gap, recent work has explored the use of LLMs in other stages of the design workflow, such as organization and verification. Works such as ROME~\cite{rome} and HiveGen~\cite{hivegen} propose LLM-assisted frameworks that decompose complex systems into sub-modules and generate them block-by-block, resembling the divide-and-conquer methodology adopted by human engineers. However, these approaches still rely on manually written testbenches and therefore fall short of an end-to-end workflow. Leveraging LLMs in the verification stage, AutoBench~\cite{autobench} and CorrectBench~\cite{correctbench} focus purely on testbench generation and self-correction, while MAGE~\cite{zhao2025mage} incorporates LLM-based internal testbench generation to assist RTL generation. However, their scope remains limited to single-module settings, not extending to hierarchical testbench generation for complex systems.
Overall, despite these meaningful explorations along different directions, there remains a lack of an LLM-based end-to-end hardware design workflow—one that can take high-level design requirements, automatically analyze and decompose tasks, generate individual modules, construct corresponding testbench for verification, and ultimately deliver a complete and functionally correct hardware system—which is what we try to propose in this paper. The comparison of our proposed framework against existing representative research is presented in Table \ref{tab:framework_com}.

\begin{table}[t]
\centering
\small
\setlength{\tabcolsep}{4pt}
\caption{Comparison of different workflows. Our framework achieves end-to-end automation for complex design.}
\begin{tabular}{lccc}
\toprule
Framework & \makecell{RTL\\Generation} & \makecell{Testbench\\Generation} & \makecell{Complex Design\\Divide \& Conquer} \\
\midrule
\cite{verigen,liu2024rtlcoder,pei2024betterv,rtl++,cui2024origen} & \cmark & \xmark & \xmark \\
AutoBench    & \xmark & \cmark & \xmark \\
CorrectBench & \xmark & \cmark & \xmark \\
ROME         & \cmark & \xmark & \cmark \\
HiveGen      & \cmark & \xmark & \cmark \\
MAGE         & \cmark & \cmark & \xmark \\
\midrule
\textbf{Ours} & \cmark & \cmark & \cmark \\
\bottomrule
\end{tabular}
\label{tab:framework_com}
\end{table}

In order to enhance our framework’s capability in handling complex hardware designs, we introduce two hierarchical intermediate representations (IRs) that standardize the description of hierarchical systems: the \textbf{Architectural Sketch}, which captures top-level structural organization, and the \textbf{Operational Specification}, which defines module functionalities and interfaces. These IRs serve two key purposes. First, they enable our framework to extract design requirements from natural language or diagrams and organize them in a structured form, facilitating both RTL and testbench generation. Second, they provide an additional level of abstraction for hardware design, allowing LLMs to better understand and reason about complex systems, and to design them effectively using a divide-and-conquer strategy. Building on these IRs, we propose an end-to-end multi-agent framework that redefines existing LLM-assisted hardware design workflows. Our framework begins with a planning stage, where the input is analyzed to generate the Architectural Sketch, followed by the Operational Specifications for each module.
Once the IRs are finalized, they serve as guides for the generation, where RTL and corresponding testbenches are generated block-by-block. Adopting a bottom-up approach, it first generates all the sub-modules and then integrates them into the top-module.

In the RTL generation phase, a self-reflection loop is adopted by many existing frameworks \cite{cui2024origen,zhao2025mage,rome,hivegen}, where simulation outputs are fed back to the generator LLM for iterative debug. However, we observe that such a vanilla debug loop is insufficient for complex modules, particularly for sub-modules with intricate logic or the top-modules. To address this limitation, we design an effective multi-agent debug loop within our framework. Specifically, the loop consists of: (1) a probing information generator that identifies relevant internal signals to monitor, (2) an automated script that instruments the testbench to capture these signals, (3) a debug advisor that analyzes the execution traces and provides diagnostic advice, and (4) an error fixer that updates the RTL based on the generated advice. In addition, to prevent repetitive errors across iterations, we maintain a debug history that records the error patterns, corresponding advice, and RTL modifications.
With this multi-agent debug loop, our framework is able to reliably produce functionally correct designs for complex systems.


We first evaluate our framework on the widely used Verilog-Eval benchmark\cite{liu2023verilogeval}. Compared to the existing state-of-the-art baseline\cite{zhao2025mage}, our approach achieves a higher pass rate, demonstrating the effectiveness of the proposed IRs. However, as our goal is to enable complex, multi-module system design, Verilog-Eval does not fully capture the capabilities of our framework. To address this, we construct a case study consisting of seven representative complex hardware systems, ranging from digital signal processing (e.g., a beamformer) to processor designs (e.g., a RISC-V processor), along with 29 extracted sub-modules. For fair evaluation, we manually develop golden testbenches for all cases.
We compare our framework with ROME\cite{rome} and MAGE\cite{zhao2025mage} on this case study. While ROME and MAGE achieve only 0/7 and 1/7 success rates on full-system designs, and 16/29 and 19/29 on sub-modules, respectively, our framework successfully passes all cases. These results highlight the effectiveness of our approach in enabling end-to-end design of complex hardware systems. 


\section{The Intermediate Representations}
A typical RTL design methodology proceeds by progressively reducing abstraction from a high‑level architectural description to an implementation‑ready logic design. The process often begins by defining the design hierarchy and interconnection among major blocks to form an abstract blueprint. The structural definition is then followed by decomposing the overall system by operation into the functional responsibilities of individual blocks. The final stage describes the logical behavior of each block, including its internal data flow, control, and timing.

While LLMs can often micro‑architect relatively small, simple modules by inferring data flow and control from a concise prompt, they are less reliable when handling hierarchical designs. Such designs require accurate connectivity descriptions, precise port mapping, and operational partitioning across blocks— information that can be difficult for an LLM to infer reliably. To mitigate this limitation, we propose two structured intermediate representations(IRs). The first IR - the Architectural Sketch captures the high‑level hierarchy and interconnection among major blocks, while the subsequent Operational Specification partitions overall functionality across constituent sub-modules. These IRs are generated by two LLM agents in our framework, the Architecture Designer and the Function Decomposer. Together, they pave an automated structured pathway from design intent to implementation‑ready RTL.

\subsection{Architectural Sketch}


One of the major challenges in hardware design lies in architectural organization.
For relatively simple tasks, such as single-module hardware designs that previous studies have primarily focused on, current LLMs may be capable of generating reasonable code from textual descriptions. However, for complex systems composed of multiple interacting modules, the interconnection and control among modules introduce intricate timing and even state transitions, posing a major challenge for LLM-aided hardware design. Under such circumstances, providing detailed textual descriptions of the system architecture not only increases the burden on designers, but also leads to longer prompts that may cause the LLM to hallucinate or fail to accurately capture certain design details due to the limited context window length. 

To systematically represent the architectural organization, we introduce the Architectural Sketch, which is extracted by the LLM Architecture Designer.
This JSON-written sketch constitutes a structural representation of the RTL design, situated between a high‑level textual description and a fully detailed behavioral implementation. At this abstraction level, the IR captures the module hierarchy, instance names, and data connectivity, including explicit signal names and port‑width dimensions derived from the port headers.

We only specify flow of data and interconnection at this stage, rather than top-level control logic or individual block level functionality.
For the data path, a uniform handshaking protocol is designed, with each block exposing a valid input signal (in\_valid) and a valid output signal (out\_valid) that together define the data interfaces along the pipeline. Overall, this IR corresponds to the structural integration or top‑level wiring stage of a general RTL design pipeline, conveying structural information rather than control‑logic behavior or operation‑specific semantics.


\begin{figure}[h]
    \centering
    \includegraphics[width=\linewidth]{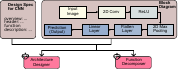}
    \caption{Inputs of the CNN Accelerator: Design Spec and Block Diagram. This is an actual case included in our case study.}
    \label{fig:CNN}
\end{figure}

We take an example of a Convolutional Neural Network (CNN) Accelerator from our proposed case study to describe the IR generation. As shown in Figure \ref{fig:CNN}, the only inputs to our end-to-end workflow are a block diagram and a design spec with specific details such as input image size, intermediate tensor size, kernel size. The diagram and prompt are first used to create the Sketch by the Architecture Designer. The Sketch highlighted by Figure ~\ref{fig:cnn_arch} specifies five sub-modules---Conv2D, ReLU, Maxpool2D, Flatten, and Linear--- all connected sequentially. From the sketch it is evident that the input \texttt{pixel\_in} signal feeds Conv2D, whose \texttt{data\_out} drives ReLU. The Maxpool2D block then expands the 16-element input from ReLU to a 32-element \texttt{data\_out}, which flows through Flatten to Linear, producing the 3-class classification \texttt{result}. The input \texttt{weight\_in} signals connect to both Conv2D (\texttt{w\_in}) and Linear (\texttt{w\_in}). The  \texttt{clk} and \texttt{rst\_n} are common to all blocks, while the \texttt{in\_valid} and \texttt{out\_valid} handshaking signals manage the pipeline. Importantly, this Sketch conveys \emph{no functional semantics}---it neither specifies how Conv2D performs convolution nor how Maxpool2D computes maximums, nor defines the precise ReLU threshold. The JSON captures purely \emph{structural connectivity}: which ports connect to which ports, module instantiations, and port widths---leaving algorithmic details and control implementation to the next IR. This architectural sketch can translate textual descriptions of module connectivity and/or a diagram shown in Figure~\ref{fig:CNN} into a formal representation of interconnection, still at an abstraction above pure RTL. 

\begin{figure}[h]
\centering
\setlength{\abovecaptionskip}{2pt}
\begin{tcolorbox}[colback=white, colframe=black, boxrule=0.5pt, arc=2pt,
                  left=4pt,right=4pt,top=2pt,bottom=2pt]
\scriptsize\ttfamily
\setlength{\parindent}{0pt}
\newcommand{\jline}[2]{\hangindent=#1\hangafter=1\hspace*{#1}#2\par}
\jline{0em}{\textcolor{aesTeal}{"inputs"}: [}
\jline{1em}{\textcolor{aesCoral}{"pixel\_in[0..15]"},}
\jline{1em}{\textcolor{aesCoral}{"weight\_in[0..15]"},}
\jline{1em}{\textcolor{aesCoral}{"clk"}, \textcolor{aesCoral}{"rst\_n"}, \textcolor{aesCoral}{"cmd"}, \textcolor{aesCoral}{"in\_valid"}}
\jline{0em}{],}
\jline{0em}{\textcolor{aesTeal}{"outputs"}: [}
\jline{1em}{\textcolor{aesCoral}{"result[0..2]"}, \textcolor{aesCoral}{"out\_valid"}}
\jline{0em}{],}
\jline{0em}{\textcolor{aesTeal}{"modules"}: \{}
\jline{1em}{\textcolor{aesBlue}{\textbf{"Conv2D"}}: \textcolor{aesCoral}{"conv"},}
\jline{1em}{\textcolor{aesBlue}{\textbf{"ReLU"}}: \textcolor{aesCoral}{"relu"},}
\jline{1em}{\textcolor{aesBlue}{\textbf{"Maxpool2D"}}: \textcolor{aesCoral}{"pool"},}
\jline{1em}{\textcolor{aesBlue}{\textbf{"Flatten"}}: \textcolor{aesCoral}{"flat"},}
\jline{1em}{\textcolor{aesBlue}{\textbf{"Linear"}}: \textcolor{aesCoral}{"fc"}}
\jline{0em}{\},}
\jline{0em}{\textcolor{aesTeal}{"Data Connections"}: \{}
\jline{1em}{\textcolor{aesPurple}{"pixel\_in[i] where i=0..15"}: [\textcolor{aesCoral}{"conv.pix\_in[i]"}],}
\jline{1em}{\textcolor{aesPurple}{"weight\_in[i] where i=0..15"}: [\textcolor{aesCoral}{"conv.w\_in[i]"}, \textcolor{aesCoral}{"fc.w\_in[i]"}],}
\jline{1em}{\textcolor{aesPurple}{"conv.data\_out[j] where j=0..15"}: [\textcolor{aesCoral}{"relu.data\_in[j]"}],}
\jline{1em}{\textcolor{aesPurple}{"relu.data\_out[j] where j=0..15"}: [\textcolor{aesCoral}{"pool.data\_in[j]"}],}
\jline{1em}{\textcolor{aesPurple}{"pool.data\_out[m] where m=0..31"}: [\textcolor{aesCoral}{"flat.data\_in[m]"}],}
\jline{1em}{\textcolor{aesPurple}{"flat.data\_out[m] where m=0..31"}: [\textcolor{aesCoral}{"fc.data\_in[m]"}],}
\jline{1em}{\textcolor{aesPurple}{"fc.scores\_out[k] where k=0..2"}: [\textcolor{aesCoral}{"result[k]"}]}
\jline{0em}{\},}
\jline{0em}{\textcolor{aesTeal}{"headers"}: \{}
\jline{1em}{\textcolor{aesBlue}{\textbf{"Conv2D"}}: \textcolor{aesCoral}{"module Conv2D(clk, rst\_n, cmd, in\_valid, pix\_in[0:15], w\_in[0:15], data\_out[0:15], out\_valid)"},}
\jline{1em}{\textcolor{aesBlue}{\textbf{"ReLU"}}: \textcolor{aesCoral}{"..."},}
\jline{1em}{\textcolor{aesBlue}{\textbf{"Maxpool2D"}}: \textcolor{aesCoral}{"..."},}
\jline{1em}{\textcolor{aesBlue}{\textbf{"Flatten"}}: \textcolor{aesCoral}{"..."},}
\jline{1em}{\textcolor{aesBlue}{\textbf{"Linear"}}: \textcolor{aesCoral}{"..."},}
\jline{1em}{\textcolor{aesBlue}{\textbf{"top\_module"}}: \textcolor{aesCoral}{"module top\_module(clk, rst\_n, cmd, in\_valid, pixel\_in[0:15], weight\_in[0:15], out\_valid, result[0:2])"}}
\jline{0em}{\}}
\end{tcolorbox}
\caption{Arch Sketch of the CNN Accelerator}
\label{fig:cnn_arch}
\end{figure}
Leveraging this IR, our framework achieves higher automation as human designers provide only high-level inputs with block diagrams and/or natural language prompts, from which the system automatically generates the Architectural Sketch. This eliminates the need of manual RTL wiring and port specification. 
This Sketch serves as the foundation for task decomposition performed in the next step, enabling a modular design and verification flow across individual sub-modules.

\subsection{Operational Specification} 
The natural next step after the Architectural Sketch would be formalizing functional intent into machine‑interpretable form. This purpose is fulfilled by the second IR, the Operational Specification, generated by the Function Decomposer agent. It views the hardware functionality from an operational perspective, where each operation reflects a separate part of the module function, such as configuration or computation. For each module, it lists all possible operations and for every operation, specifies the input pattern, output pattern, and the width and type of each involved port, thereby turning high‑level interface descriptions into precise data flow. It also includes a textual description of the operation that captures the semantics of the operation. Importantly, this IR organizes the specification hierarchically. The top‑module operations are decomposed into the corresponding sub‑operations for sub-module instances, thus propagating control and data flow down the design hierarchy. In this way, the Operational Specification IR systematically converts abstract architecture and natural‑language‑style prompts into a detailed, structured description of actual functionality, bridging the gap between architectural intent and the RTL implementation. The Operational Specification is structured in the following way:
\begin{itemize}[leftmargin=*]
    \item \textbf{Module header}: Specifies the complete Verilog port interface
    \item \textbf{Exhaustive operation list} covering every supported operation:
    \begin{itemize}[leftmargin=0.8em]
        \item \textbf{Name}: Operation identifier
        \item \textbf{Flow Control signals}: Exact trigger conditions for data capture and launch (e.g., \texttt{cmd=2'b11 \&\& in\_valid})
        \item \textbf{Input pattern}: Data transfer protocol indicating single cycle input, multi-beat streaming
        \item \textbf{Output pattern}: Result format and delivery mechanism
        \item \textbf{Description}: Textual functional specification
    \end{itemize}
\end{itemize}

\begin{figure}[t]
\centering
\setlength{\abovecaptionskip}{2pt}
\begin{tcolorbox}[colback=white, colframe=black, boxrule=0.5pt, arc=2pt,
                  left=4pt,right=4pt,top=2pt,bottom=2pt]
\scriptsize\ttfamily
\setlength{\parindent}{0pt}
\newcommand{\jline}[2]{\hangindent=#1\hangafter=1\hspace*{#1}#2\par}
\jline{0em}{\textcolor{aesBlue}{\textbf{"Conv2D"}}: \{}
\jline{1em}{\textcolor{aesTeal}{"header"}: \textcolor{aesCoral}{"..."},}
\jline{1em}{\textcolor{aesTeal}{"operations"}: [\{}
\jline{2em}{\textcolor{aesTeal}{"name"}: \textcolor{aesCoral}{"LOADW"},}
\jline{2em}{\textcolor{aesTeal}{"control\_signals"}: [\textcolor{aesCoral}{"cmd == 2'b01"}, \textcolor{aesCoral}{"in\_valid"}],}
\jline{2em}{\textcolor{aesTeal}{"input\_pattern"}: \{}
\jline{3em}{\textcolor{aesPurple}{"type"}: \textcolor{aesCoral}{"streaming"},}
\jline{3em}{\textcolor{aesPurple}{"beats"}: \textcolor{aesAmber}{2},}
\jline{3em}{\textcolor{aesPurple}{"data\_ports"}: [\{\textcolor{aesPurple}{"name"}: \textcolor{aesCoral}{"w\_in"}, \textcolor{aesPurple}{"width"}: \textcolor{aesAmber}{16}, \textcolor{aesPurple}{"count"}: \textcolor{aesAmber}{16}\}],}
\jline{3em}{\textcolor{aesPurple}{"protocol"}: \textcolor{aesCoral}{"cmd == 2'b01 and in\_valid high for 2 cycles. w\_in[0:15] carries 16 int16 values per beat, for a total of 32 values."}}
\jline{2em}{\},}
\jline{2em}{\textcolor{aesTeal}{"output\_pattern"}: \{ \textcolor{aesPurple}{"type"}: \textcolor{aesCoral}{"none"} \},}
\jline{2em}{\textcolor{aesPurple}{"description"}: \textcolor{aesCoral}{"Load 3x3 kernels and biases for 2 output channels."}}
\jline{1em}{\}]}
\jline{0em}{\},}
\jline{0em}{\ }
\jline{0em}{\textcolor{aesBlue}{\textbf{"ReLU"}}: \{ ... \},}
\jline{0em}{\textcolor{aesBlue}{\textbf{"Maxpool2D"}}: \{ ... \},}
\jline{0em}{\textcolor{aesBlue}{\textbf{"Flatten"}}: \{ ... \},}
\jline{0em}{\textcolor{aesBlue}{\textbf{"Linear"}}: \{ ... \},}
\jline{0em}{\ }
\jline{0em}{\textcolor{aesBlue}{\textbf{"top\_module"}}: \{}
\jline{1em}{\textcolor{aesTeal}{"header"}: \textcolor{aesCoral}{"..."},}
\jline{1em}{\textcolor{aesTeal}{"operations"}: [\{}
\jline{2em}{\textcolor{aesTeal}{"name"}: \textcolor{aesCoral}{"RUN"},}
\jline{2em}{\textcolor{aesTeal}{"control\_signals"}: [\textcolor{aesCoral}{"cmd == 2'b11"}, \textcolor{aesCoral}{"in\_valid"}, \textcolor{aesCoral}{"out\_valid"}],}
\jline{2em}{\textcolor{aesTeal}{"input\_pattern"}: \{}
\jline{3em}{\textcolor{aesPurple}{"type"}: \textcolor{aesCoral}{"streaming"},}
\jline{3em}{\textcolor{aesPurple}{"beats"}: \textcolor{aesAmber}{4},}
\jline{3em}{\textcolor{aesPurple}{"data\_ports"}: [\{\textcolor{aesPurple}{"name"}: \textcolor{aesCoral}{"pixel\_in"}, \textcolor{aesPurple}{"width"}: \textcolor{aesAmber}{16}, \textcolor{aesPurple}{"count"}: \textcolor{aesAmber}{16}\}],}
\jline{3em}{\textcolor{aesPurple}{"protocol"}: \textcolor{aesCoral}{"cmd == 2'b11 and in\_valid 4 consecutive cycles to deliver the image"}}
\jline{2em}{\},}
\jline{2em}{\textcolor{aesTeal}{"output\_pattern"}: \{}
\jline{3em}{\textcolor{aesPurple}{"type"}: \textcolor{aesCoral}{"single"},}
\jline{3em}{\textcolor{aesPurple}{"beats"}: \textcolor{aesAmber}{1},}
\jline{3em}{\textcolor{aesPurple}{"data\_ports"}: [\{\textcolor{aesPurple}{"name"}: \textcolor{aesCoral}{"result"}, \textcolor{aesPurple}{"width"}: \textcolor{aesAmber}{16}, \textcolor{aesPurple}{"count"}: \textcolor{aesAmber}{3}\}],}
\jline{3em}{\textcolor{aesPurple}{"protocol"}: \textcolor{aesCoral}{"out\_valid pulses high for 1 cycle with the single output beat carrying 3 classification scores."}}
\jline{2em}{\},}
\jline{2em}{\textcolor{aesPurple}{"description"}: \textcolor{aesCoral}{"Run the CNN inference pipeline: Conv2D -> ReLU -> Maxpool2D -> Flatten -> Linear (Q8.8 fixed-point)."}}
\jline{1em}{\}]}
\jline{0em}{\}}
\end{tcolorbox}
\caption{Example of Operational Specification of a CNN}
\label{fig:json_spec}
\end{figure}

Figure 3 presents a JSON-based operational specification for the CNN accelerator, illustrating how module behavior, interfaces, and control protocols are captured in a machine-readable yet human-interpretable form. For clarity and space efficiency, we show only a single representative operation for both the Conv2D block and the top-level module, and omit the full operation sets of the intermediate sub-modules (ReLU, Maxpool2D, Flatten, and Linear), which follow the same structural pattern. The included operations focus on the end-to-end path, specifying the streaming input protocol, output behavior, and inter-module control flow. Descriptions are intentionally trimmed to convey intent without reproducing the full operational semantics, which are defined elsewhere in the specification. This excerpt is sufficient to demonstrate how the JSON format concisely encodes hardware operation behavior, data movement, and module orchestration across the CNN inference pipeline.This IR's utility spans design paradigms from combinational logic, which can be described with a single operation, to FSM controllers chaining multiple operations in a particular sequence, to complete processor instruction sets and dataflow accelerators.

The Operational Specification forms an important contribution of this work. It replaces high-level human reasoning, requiring deep engineering judgment with structured, executable behavioral contracts that LLMs can reliably interpret and implement. By providing exhaustive information on control, data patterns, and specific functional descriptions, the IR eliminates ambiguity, enabling LLMs to bridge the critical gap of semantic disconnect between high-level system architecture and module-level implementation.

\section{The Multi-Agent Framework}
\begin{figure}[h]
    \centering
    \includegraphics[width=.9\linewidth]{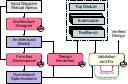}
    \caption{Overall workflow of the multi-agent framework.}
    \label{fig:framework}
\end{figure}

As depicted in Figure \ref{fig:framework}, the proposed multi-agent framework is divided into two major stages: the \textbf{Planning stage}, which converts diverse forms of input into detailed IRs, i.e. the \textbf{Architectural Sketch} and \textbf{Operational Specification}, and the \textbf{Generation Stage}, which automates the generation of both the RTL and testbench, verification, and iterative debug with a multi-agent debug loop.

\subsection{Planning Stage}
Similar to the human design process, the framework takes a block diagram and a textual specification as input. It first performs system-level analysis, constructing the Architectural Sketch with the Architecture Designer Agent, which determines the sub-modules and their connections. With this data, the Function Decomposer Agent generates the Operational Specifications for all sub-modules and the top module. 
It also produces a software function for each module, specifying the algorithm of \textit{output = F(input)} for each operation corresponding to the Operational Specification. 
The software functions are crucial to build the hierarchical testbenches for the following verification process. Note that no human intervention is needed during the end-to-end workflow.


\subsection{Generation Stage}
Once the IRs are generated, the planning stage concludes and the framework transitions into the generation stage. In this stage, the Architectural Sketch and Operational Specifications are read by the Design Generator agent to schedule the design tasks for each block. Sub-module generation proceeds independently, requiring only the corresponding Operational Specification to produce and verify the RTL implementation. Upon successful verification of all sub-modules, the top-module is constructed using the Architectural Sketch, its Operational Specification, and the generated sub-module RTLs. Together, these provide the information necessary to orchestrate top-level control. This bottom-up approach ensures that each level of the hierarchy is verified before proceeding to the next.

A key challenge for end-to-end RTL generation frameworks lies in the sourcing of testbenches. Prior work such as ROME~\cite{rome} and HiveGen~\cite{hivegen} relies on manually written testbenches or those provided by existing benchmarks, limiting scalability. More recent efforts including AutoBench~\cite{autobench}, CorrectBench~\cite{correctbench}, and MAGE~\cite{zhao2025mage} explore LLM-based testbench generation, but their scope is confined to single-module complexity and does not extend to hierarchical designs. Moreover, purely LLM-generated testbenches often suffer from limited reliability. In some cases, they diverge from the intended design semantics or impose constraints that are either overly stringent or inconsistent with the specification.

To address this challenge, we explicitly account for the requirements of automated testbench generation in the design of the Operational Specification IR. For complex designs, we decompose the testbench into two key components: the input/output flow patterns and the reference function, i.e., \textit{output = F(input)}. The former is captured in the Operational Specification, which prescribes the input trigger and output sampling conditions for each operation, along with the cycle-level behavior of inputs and outputs. The latter is obtained directly from the software function generated alongside the Operational Specification by the Function Decomposer. Together, these two components allow the framework to construct testbenches deterministically,
without relying purely on LLMs at the verification stage, yielding higher confidence in correctness. The testbench extracts the input/output patterns from the Operational Specification, issues multiple sets of randomized stimuli, and samples the corresponding outputs. The same stimuli are then applied to the software reference function, and the outputs of the hardware and software models are compared to validate functional equivalence. The testbench performs verification on a per-operation basis, where each operation is associated with a corresponding software reference function. In this way, our framework naturally supports the generation of hierarchical testbenches. The top-level testbench only needs to invoke the top-level software reference function, which in turn calls the corresponding software functions of sub-modules, thereby ensuring functional consistency across different levels of verification. In this way, our framework achieves fully end-to-end hardware design without the need for external testbenches.

\subsection{Multi-Agent Debug Loop}
\begin{figure}[h]
    \centering
    
    \begin{subfigure}{\linewidth}
        \centering
        \includegraphics[width=0.6\linewidth]{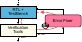}
        \caption{Vanilla self-reflection debug loop.}
        \label{fig:debug_comparea}
    \end{subfigure}
    
    \vspace{0.5em}
    
    \begin{subfigure}{\linewidth}
        \centering
        \includegraphics[width=.95\linewidth]{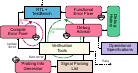}
        \caption{Proposed multi-agent debug loop.}
        \label{fig:debug_compareb}
    \end{subfigure}
    
    \caption{Comparison between vanilla self-reflection and our proposed multi-agent debug loop.}
    \label{fig:debug_compare}
\end{figure}

Self-reflection loops have been adopted in many prior research\cite{cui2024origen,zhao2025mage,rome,hivegen}, where error feedback after verification is fed back to the LLM for iterative debug. However, after experimenting with a vanilla debug loop (shown in Figure \ref{fig:debug_comparea}), we found that it is insufficient to meet the demands of complex design tasks, where some modules, such as the top-modules, may involve intricate logic. We identify two key limitations of the basic debug loop. First, it relies solely on output mismatches for debug, without access to internal signals. For complex hierarchical systems, such end-level mismatches often fail to provide sufficient information to diagnose errors. Second, each debug iteration is conducted as an independent interaction, which makes it repeatedly introduce the same errors across iterations.

To address these limitations, we extend the basic self-reflection loop by introducing an enhanced multi-agent debug loop, which is presented in Figure \ref{fig:debug_compareb}. 
As shown in the figure, when a module fails verification, the resulting errors are classified as syntax or functional errors. Syntax errors are handled by sending the module and error back directly to the generator LLM for regeneration. Functional errors are handled through a multi-agent debug loop. The framework first invokes a probe LLM to identify the key internal signals that require monitoring, mirroring the practice of human engineers, who select a small set of critical signals among numerous waveform traces. The testbench is then instructed to observe the selected signals, and the RTL is executed again under this testbench, with signal values recorded on a cycle-by-cycle basis. To avoid excessive verbosity, values are logged only on transitions. The resulting trace, together with the corresponding IR, the original RTL, and the error information, is provided to an analyzer LLM that diagnoses the root cause and produces targeted debug advice. This advice is subsequently passed to a fixer LLM, which revises the design and produces an improved RTL implementation. To prevent the framework from repeating prior mistakes, we maintain a history that tracks the error, debug advice, and RTL difference at each iteration. Collectively, this multi-agent debug loop substantially improves debug effectiveness on complex system-level designs compared to basic self-reflection loops, as we demonstrate in our ablation study.


\section{Evaluation}
For evaluation, we first review existing frameworks and assess our approach on the widely adopted Verilog-Eval benchmark. While the results demonstrate the effectiveness of our framework, this benchmark does not fully reflect our capabilities, as the benchmark consists solely of small, single-module designs. To better evaluate performance on complex, hierarchical systems, we introduce a new case study comprising seven representative complex hardware designs and 29 extracted sub-modules. We conduct experiments on this case study, comparing our framework against two representative baselines, ROME~\cite{rome} and MAGE~\cite{zhao2025mage}, and perform synthesis on the generated RTL. Finally, we conduct ablation studies to evaluate the contributions of the two IRs and the multi-agent debug loop.

We evaluate the framework using four LLMs. Their API model identifiers are GPT-5 (\nolinkurl{gpt-5-2025-08-07}), GPT-5.4 (\nolinkurl{gpt-5.4}), Claude Sonnet (\nolinkurl{claude-sonnet-4-6}), and Claude Opus (\nolinkurl{claude-opus-4-6}). We synthesize the generated RTL using Synopsys Design Compiler with the GlobalFoundries 22FDX 22\,nm standard-cell library and report cell counts from the resulting netlists.

\subsection{Experiment on Existing Benchmark}
For Verilog-Eval assessment, we use the code-complete-iccad2023 dataset. For our framework, the input prompt is first processed to generate an Operational Specification, which is then provided to the LLM (GPT-5) together with the prompt. The Architectural Sketch is omitted, as the benchmark consists of small, single-module designs that do not require organization. In addition, we disable testbench generation and the debug loop in this setting. Therefore, this experiment primarily aims to evaluate whether the Operational Specification can effectively improve RTL generation quality. We compare the result with a native GPT-5 generation baseline. In addition, we use GPT-5 to rerun the open-source state-of-the-art framework MAGE, which features internal testbench generation and iterative debug using a multi-agent approach. For each setup, we perform 5 independent runs and get the Pass@5 result.
\begin{table}[h]
\caption{Verilog-Eval Results Comparison. Pass@5(clean) refers to the result excluding the 7 cases with issues.}
\label{tab:compare}
\centering
\small
\begin{tabular}{lccc}
\toprule
\textbf{Method}  &\textbf{Pass@5} &  \textbf{v.s. Native}& \textbf{Pass@5(clean)}  \\
\midrule
Native GPT-5 & 146/156 & 0 & 146/149  \\
MAGE(with GPT-5) & 146/156 & 0 & 144/149  \\
\textbf{Ours(with GPT-5)} & 149/156 & \color{ForestGreen}+3 & 149/149  \\
\bottomrule
\end{tabular}
\end{table}

As shown in Table~2, both native GPT-5 generation and MAGE pass 146 out of 156 cases, failing on 10, whereas our framework increases the pass count to 149, reducing failures to 7. Notably, the failures of our framework form a strict subset of those from the native baseline, indicating consistent improvement. These results suggest that augmenting prompts with the Operational Specification enhances single-module RTL generation quality. Upon further analysis, we observe that Verilog-Eval contains several problems with issues (IDs 53, 62, 93, 98, 104, 133, and 149), including incorrect reset handling, synthesizability constraints, and prompt ambiguity, as also noted in~\cite{hsin2026evolve}. For example, in Problem 62, the golden model and testbench interpret \texttt{sel} wrongly from the prompt, while in Problem 93, the reference behavior itself is functionally incorrect. Such ambiguity and inconsistencies are more prevalent in the latest V2 dataset, which we therefore do not adopt for evaluation. After excluding these cases, our framework achieves a 100\% pass rate.

\subsection{Experiments on Proposed Case Study}
While Verilog-Eval provides a useful foundation for evaluation, it primarily targets simple, flat hardware blocks with limited structural complexity. It does not fully capture the challenges that arise when reasoning over nested sub-module instantiations, complex control logic, or programmable data paths — all of which are routine in real-world hardware systems. To comprehensively evaluate the capability of LLM-assisted frameworks on realistic hardware generation tasks, complex and hierarchical designs are necessary. Therefore, we propose a case study consisting of seven complex hardware designs (Beamformer, OFDM, Image Match, CNN, Systolic Array, Vector Unit and RISC-V processor), and 29 sub-modules extracted from these designs. The input to the CNN case is shown in Figure \ref{fig:CNN} as an example. The case study is designed from the ground up to include complex module hierarchies, non-trivial control and datapath logic, and multi-module datapaths that test both structural and functional correctness. We manually created golden testbenches as a final check to ensure that the generated testbenches are functionally correct. Note that, this is just for validation and is not needed for our framework itself, as it will generate the testbench in the end-to-end workflow.

\begin{table}[h]
\caption{RTL generation statistics of our cases using our framework, compared to Verilog-Eval cases, all using GPT-5.4. Sub refers to sub-module number. LOC refers to the total lines of RTL code for each case.}
\label{tab:detail}
\centering
\setlength{\tabcolsep}{4pt}
\begin{tabular}{@{}l@{\hspace{3pt}}ccccc@{}}
\toprule
Case & Time(s) & Tokens & Sub & LOC & Cell Count \\
\midrule
Beamformer        & 576  & 84K  & 5 & 925  & 6.3M \\
OFDM              & 447  & 34K  & 5 & 558  & 8K \\
Image Match       & 538  & 53K  & 4 & 622  & 58K \\
CNN Accelerator   & 239  & 29K  & 5 & 578  & 79K \\
Systolic Array    & 354  & 31K  & 4 & 537  & 7K \\
Vector Unit       & 2641 & 194K & 2 & 1709 & 3M \\
RISC-V Processor  & 802  & 148K & 4 & 514  & 22K \\
\midrule
\textbf{Average}  & \textbf{$\sim$800} & \textbf{$\sim$82K} & \textbf{$\sim$4} & \textbf{$\sim$780} & \textbf{$\sim$1.4M} \\
\midrule
Veri-Eval Maximum  & 10 & 1.5K     & -- & 65 & 5.4K \\
Veri-Eval Average  & 3 & 0.377K & -- & 16 & 101 \\
\bottomrule
\end{tabular}
\end{table}

The seven designs are selected to collectively span the major domains of modern digital hardware design. The Beamformer and OFDM address signal processing and communications in wireless and radar systems; the Image Match unit represents computer vision workloads; the CNN accelerator and Systolic Array represent the computation paradigms in machine learning acceleration; and the RISC-V Processor and Vector Unit represent general-purpose and data-parallel programmable architectures, respectively. Together, these designs are deliberately chosen such that they span the full landscape of application domains. Most hardware design tasks are likely to share structural or computational characteristics with at least one module in this set. 

In terms of RTL design complexity, the modules stress-test different dimensions of LLM-based design capability. The RISC-V Processor and Vector Unit challenge the model with programmability. The large number of instructions, opcodes, and control flow cases that must all be correctly accounted for is not easy to handle. The OFDM is an intricate FSM with complex state transitions, where a single wrong transition in between can cause errors. The systolic array poses more of a structural challenge, requiring correct instantiation and interconnection of a large 2D mesh of processing elements. The CNN engine and beamformer involve complex functionality and arithmetic while the image match core combines spatial search logic with arithmetic comparison across a large space. Together, these modules ensure the case study stresses the full spectrum of RTL generation challenges an LLM is likely to encounter.

Table 3 lists the statistics of successful results generated by our framework (using GPT-5.4). Tokens refer to the sum of input and output tokens used for the end-to-end generation. We synthesize the generated RTL to obtain the cell counts. The data highlights the substantial difference in complexity between our cases and Verilog-Eval. On average, our cases require 80 times more execution time, consume approximately 55 times more tokens, contain 12 times more lines of code, and include nearly 260 times more cells than the largest Verilog-Eval case. This also indicates that our case study is more comprehensive, covering substantially larger and more intricate designs.

We present our evaluation in Tables \ref{tab:benchmark} and \ref{tab:subbenchmark} comparing our framework with MAGE and ROME across the proposed case study. To enable a deeper analysis, we further examine the performance of all frameworks in generating the sub-modules corresponding to each case. All top- and sub-modules generated by MAGE, ROME, and our framework are verified against the manually written golden testbenches provided by our case study. Additionally, we evaluate its effectiveness when leveraging different state-of-the-art LLMs for RTL generation. 
As highlighted in Table \ref{tab:benchmark}, our framework passes all 7 cases using GPT-5 and GPT-5.4. ROME is not able to pass any and MAGE only passes one. In addition, our evaluation upon different LLMs (GPT-5, GPT-5.4, Claude Sonnet and Claude Opus) shows the generality of our framework, as all of them pass at least five cases.

\begin{table}[h]
\caption{Case Study results across frameworks and across LLMs. ROME and MAGE are evaluated using GPT-5.4.}
\label{tab:benchmark}
\centering
\small
\setlength{\tabcolsep}{5pt}
\resizebox{\linewidth}{!}{
\begin{tabular}{lcccccc}
\toprule
& \multicolumn{3}{c}{\textbf{Frameworks (w/ GPT-5.4)}} & \multicolumn{3}{c}{\textbf{Ours across LLMs}} \\
\cmidrule(lr){2-4} \cmidrule(lr){5-7}
\textbf{Case} & \textbf{ROME} & \textbf{MAGE} & \textbf{Ours} & \textbf{GPT-5} & \textbf{Sonnet} & \textbf{Opus} \\
\midrule
Beamformer       & \xmark & \xmark & \cmark & \cmark & \cmark & \cmark \\
CNN              & \xmark & \cmark & \cmark & \cmark & \cmark & \cmark \\
Image Match      & \xmark & \xmark & \cmark & \cmark & \cmark & \cmark \\
OFDM             & \xmark & \xmark & \cmark & \cmark & \cmark & \cmark \\
Systolic Array   & \xmark & \xmark & \cmark & \cmark & \cmark & \cmark \\
RISC-V Proc.     & \xmark & \xmark & \cmark & \cmark & \xmark & \cmark \\
Vector Unit      & \xmark & \xmark & \cmark & \cmark & \xmark & \xmark \\
\midrule
\textbf{Total} & \textcolor{red}{\textbf{0/7}} & \textcolor{red}{\textbf{1/7}} & \textcolor{ForestGreen}{\textbf{7/7}} & \textcolor{ForestGreen}{\textbf{7/7}} & \textcolor{blue}{\textbf{5/7}} & \textcolor{blue}{\textbf{6/7}} \\
\bottomrule
\end{tabular}
}
\end{table}

Though ROME claims to implement hierarchical RTL, it has one key limitation. It extracts a set of sub-modules required to implement a design from its natural language specification. For each sub-module, it expects the user to provide a corresponding unit test. If a test is available, ROME executes a debug loop using the generated RTL and the provided test; otherwise, the generated sub-module is accepted as-is. Consequently, if any sub-module within a case contains an error, ROME fails that case. 
This is also evident in Tables~\ref{tab:benchmark} and~\ref{tab:subbenchmark}, where ROME fails to correctly generate all sub-modules for any single case except the image match. Even if 
it generates the sub-modules correctly, it fails to stitch them together properly, resulting in a functional issue at the top level. This study also indicates that on some occasions ROME fails to generate correct RTL even for non-hierarchical designs, despite being provided with the unit test in the prompt. This occurs because ROME only feeds the test output back to the LLM with a request to fix the error, without re-reading the prompt at every debug iteration. 
Consequently, if the prompt is misunderstood initially, the generated RTL can never become functionally correct. This is the reason why it fails all our top-level cases and only passes 16 out of 29 simpler sub-module level tests.


\begin{table}[h]
\caption{Sub-module Case Study Results Across Frameworks}
\label{tab:subbenchmark}
\centering
\small
\setlength{\tabcolsep}{5pt}
\renewcommand{\arraystretch}{1.05}
\resizebox{\linewidth}{!}{
\begin{tabular*}{\columnwidth}{@{\extracolsep{\fill}}lccc}
\toprule
\textbf{Case} & \textbf{ROME} & \textbf{MAGE} & \textbf{Ours} \\
\midrule
Beamformer & 3/5 & 5/5 & 5/5 \\
CNN & 2/5 &2/5 & 5/5 \\
Image Match & 4/4 & 4/4 & 4/4 \\
OFDM & 3/5 & 4/5 & 5/5 \\
Systolic Array & 2/4 & 2/4 & 4/4 \\
RISC-V Processor & 2/4 & 2/4 & 4/4 \\
Vector Unit & 0/2 & 0/2 & 2/2 \\
\midrule
\textbf{Total} & \textcolor{blue}{\textbf{16/29}} & \textcolor{blue}{\textbf{19/29}} & \textcolor{ForestGreen}{\textbf{29/29}} \\
\textbf{Pass Rate} & \textcolor{blue}{\textbf{55.2\%}} & \textcolor{blue}{\textbf{65.5\%}} & \textcolor{ForestGreen}{\textbf{100\%}} \\
\bottomrule
\end{tabular*}
}
\end{table}

MAGE passes the CNN case in our case study, but fails the remaining six. Its framework generates testbenches directly from natural language to produce "State Checkpoints" used for debug, then generates initial RTL code that is evaluated against this testbench. If the RTL fails, a judge agent decides whether the testbench itself is faulty and regenerates it. If the RTL passes, a high-temperature sampling and scoring process ranks multiple RTL candidates. A potential concern with this approach is that when the RTL and testbench disagree, the judge agent must decide which to trust, risking either the rejection of correct RTL or the weakening of the testbench to match buggy code. This could lead to both false positives and false negatives. Another concern is that the debug is only as reliable as the initially generated testbench, so any upstream error can mislead the debug iterations. MAGE performs better than ROME on our case study, as it is able to pass the CNN top-module case along with 19 sub-module cases. However, unlike ROME and our framework, MAGE does not support hierarchical design. This limitation becomes more significant as top-module complexity increases, since modularity becomes necessary for scalable design. Its success on the CNN case is also consistent with Table 3, which shows that CNN is the simplest case study in terms of execution time, token count, and lines of code.

\subsection{Ablation Study}
We conduct two groups of ablation experiments. First, we compare the complete framework against relaxed-budget direct-generation and direct-repair baselines and remove each proposed IR independently. These experiments isolate whether the performance improvement arises from additional LLM calls, the Architectural Sketch, or the Operational Specification. Second, we retain the complete IR-based workflow but replace the proposed multi-agent debug loop with a vanilla self-reflection loop to evaluate the contribution of the debugging mechanism.

\subsubsection{IR and Direct-Generation Ablations}
We first compare our framework with two direct GPT-5.4 baselines using the same inputs. \textit{Direct generate-$k$} generates the RTL repeatedly and passes if any result succeeds, while \textit{direct repair-$k$} performs one generation followed by iterative debugging. The maximum budget used by our framework is 22 calls and \$1.22. To favor the direct baselines, we relax the thresholds to 25 calls and \$1.30 and terminate only after both thresholds are reached. We also evaluate the framework without each IR. \textit{w/o Op. Spec.} retains only the Architectural Sketch, while \textit{w/o Arch. Sketch} generates the system as a single module. Both ablations use at most 10 debug iterations. Table~\ref{tab:ir_ablation} presents the results.

\begin{table}[h]
\caption{Direct-generation and IR ablation results across the case
study. All configurations use GPT-5.4.}
\label{tab:ir_ablation}
\centering
\small
\setlength{\tabcolsep}{4pt}
\resizebox{\linewidth}{!}{
\begin{tabular}{@{}lccccc@{}}
\toprule
& \multicolumn{2}{c}{\textbf{Direct LLM Baselines}}
& \multicolumn{3}{c}{\textbf{Our Framework}} \\
\cmidrule(lr){2-3} \cmidrule(lr){4-6}
\textbf{Case}
& \makecell{\textbf{Direct}\\\textbf{Generate-$k$}}
& \makecell{\textbf{Direct}\\\textbf{Repair-$k$}}
& \makecell{\textbf{w/o Op.}\\\textbf{Spec.}}
& \makecell{\textbf{w/o Arch.}\\\textbf{Sketch}}
& \makecell{\textbf{Full}\\\textbf{Framework}} \\
\midrule
Beamformer
& \xmark & \xmark & \xmark & \cmark & \cmark \\
CNN
& \cmark & \cmark & \cmark & \cmark & \cmark \\
Image Match
& \xmark & \xmark & \xmark & \cmark & \cmark \\
OFDM
& \xmark & \xmark & \xmark & \xmark & \cmark \\
Systolic Array
& \xmark & \cmark & \cmark & \cmark & \cmark \\
RISC-V Processor
& \xmark & \xmark & \xmark & \cmark & \cmark \\
Vector Unit
& \xmark & \xmark & \xmark & \xmark & \cmark \\
\midrule
\textbf{Summary}
& \textcolor{red}{\textbf{1/7}}
& \textcolor{red}{\textbf{2/7}}
& \textcolor{red}{\textbf{2/7}}
& \textcolor{blue}{\textbf{5/7}}
& \textcolor{ForestGreen}{\textbf{7/7}} \\
\bottomrule
\end{tabular}
}
\end{table}

Despite the relaxed budget, direct generation and repair pass only 1/7 and 2/7 cases, respectively. Removing the Operational Specification also reduces the result to 2/7, while removing the Architectural Sketch achieves 5/7. The complete framework passes all seven cases, showing that the two IRs provide complementary functional and structural information for complex RTL generation.

\subsubsection{Multi-Agent Debug Ablation}
Furthermore, to evaluate the effectiveness of our multi-agent debug loop elaborated in Section 3.3, we conducted an ablation study. In this experiment, we reran the seven-design case study using our framework but replaced the multi-agent debug loop with a vanilla debug loop, shown in Figure \ref{fig:debug_comparea}--where only the incorrect RTL code and errors are sent back into the generation LLM for iterative debug. Compared to our multi-agent debug loop, the vanilla loop only uses mismatched and expected outputs as functional error feedback, without access to internal signal observations or historical debug information. We compare this with the original framework with the multi-agent debug loop, shown in Figure \ref{fig:debug_compareb}. Both groups use GPT-5 as the LLM, and both groups set the maximum debug iterations as 10 for each sub-module or top module. The comparison is shown in Table \ref{tab:ablation}.

\begin{table}[h]
\caption{Ablation study results across cases. MA Dbg and Van. Dbg mean our multi-agent debug loop and the vanilla debug loop used for ablation, respectively.}
\centering
\small
\setlength{\tabcolsep}{5pt}
\resizebox{\linewidth}{!}{
\begin{tabular}{@{}lcccc@{}}
\toprule
\textbf{Case} & \makecell{Sub-module\\w/ MA Dbg} & \makecell{Sub-module\\w/ Van. Dbg} & \makecell{Topmodule\\w/ MA Dbg} & \makecell{Topmodule\\w/ Van. Dbg} \\
\midrule
Beamformer       & 5/5 & 5/5 & \cmark & \cmark \\
CNN              & 5/5 & 5/5 & \cmark & \cmark \\
Image Match      & 4/4 & 4/4 & \cmark & \cmark \\
OFDM             & 5/5 & 5/5 & \cmark & \cmark \\
Systolic Array   & 4/4 & 4/4 & \cmark & \xmark \\
RISC-V Processor & 4/4 & 3/4 & \cmark & \xmark \\
Vector Unit      & 2/2 & 0/2 & \cmark & \xmark \\
\midrule
Summary          & \textcolor{ForestGreen}{29/29} & \textcolor{blue}{26/29} & \textcolor{ForestGreen}{7/7} & \textcolor{blue}{4/7}\\
\bottomrule
\end{tabular}
}
\label{tab:ablation}
\end{table}

The comparison result shows that three sub-modules fail without the multi-agent debug loop: one from RISC-V Processor and two from Vector Unit. This is expected as these cases are the most complex ones in the whole case study, and their sub-modules with relatively complicated logic would benefit the most from our multi-agent debug loop. Moreover, Systolic Array failed due to top-level error that could not be fixed by the vanilla debug loop after 10 iterations. Such errors in top modules are typically hard to debug without internal signal visibility and historical debug records, where our framework shows a clear advantage.

\section{Related Work}

Recent advances in LLMs for software programming have inspired growing interest in applying LLMs to hardware design.
One line of work focuses on improving RTL generation capability through LLM fine-tuning, where the quality of training data plays a critical role and the scarcity of high-quality hardware code remains a major challenge. Verigen~\cite{verigen} collects large-scale Verilog data from open-source repositories and textbooks, while Verilog-Eval~\cite{liu2023verilogeval} augments it with GPT 3.5-generated annotations and introduces a widely adopted benchmark. RTLCoder~\cite{liu2024rtlcoder} further generates Verilog datasets from keywords using GPT 3.5 and surpasses the base model. Subsequent works improve data quality and model performance: Origen~\cite{cui2024origen} employs dataset augmentation and self-reflection, and BetterV~\cite{pei2024betterv} uses domain-specific fine-tuning with generative discriminators, both outperforming GPT 4 on Verilog-Eval. RTL++\cite{rtl++} explores structural representations via textualized control and data flow graphs, and VerilogDB\cite{verilogdb} constructs the largest high-quality Verilog dataset to date. While these efforts demonstrate the strong potential of LLMs for RTL generation, they primarily focus on improving model performance for single-module code generation.

Several studies also investigate the capability of LLMs to support hardware design flow rather than solely RTL generation. Chip-chat~\cite{chipchat} and ChatCPU~\cite{chatcpu} present case studies on CPU design using LLMs. ROME~\cite{rome} proposes a hierarchical prompting framework to decompose complex systems into sub-modules, while HiveGen~\cite{hivegen} further leverages retrieval-augmented generation to reuse existing modules from a code library. MAGE~\cite{zhao2025mage} introduces a multi-agent framework that converts waveform information into textual feedback for debug. AutoBench~\cite{autobench} and CorrectBench~\cite{correctbench} focus on automated testbench generation from hardware descriptions. More recently, VeriOpt\cite{tasnia2025veriopt} explores PPA optimization through a multi-agent framework with iterative feedback from EDA tools. Despite these advances, existing approaches still address only parts of the hardware design workflow and remain limited in their ability to handle complex, hierarchical systems in an end-to-end manner. In particular, they either rely on manually written testbenches, focus on single-module settings, or lack coordinated generation, verification, and debug across modules. Therefore, an end-to-end framework that fully automates RTL generation, testbench creation, testing, and debug for complex hardware systems remains largely unexplored.

\section{Conclusion and Future Work}
We present an end-to-end, multi-agent framework for LLM-based hardware development, built around two hierarchical intermediate representations: the Architectural Sketch, which captures module topology and interconnection, and the Operational Specification, which defines per-module functionality. These IRs help the framework to decompose a complex design into a hierarchy of sub-modules and derive how each should be implemented and verified. The framework also leverages a powerful multi-agent debug loop to converge toward structurally and functionally correct implementations. Our framework achieves a pass@5 of 95.5\% on the widely adopted Verilog-Eval benchmark, surpassing state-of-the-art LLM-based generation baselines. Recognizing that such single-module benchmarks do not capture the complexity of real-world hardware, we introduce a new case study consisting of complex hardware designs. Our framework passes all cases and is the only evaluated framework that consistently generates structurally and functionally correct RTL for the complex designs in the case study.

In the future, we look to close a PPA (Power, Performance, Area) optimization loop on top of the proposed workflow. By incorporating feedback from physical-design tools, future versions of the framework would enable agents to reason about delay, area, and power trade-offs, iteratively refining the IRs and implementations to meet user-specified design targets. This would move the LLM-based framework toward fully automated, specification-to-silicon hardware design.



\bibliographystyle{ACM-Reference-Format}
\bibliography{refs}

\appendix









\end{document}